%% file: jocksch_parallel19.tex
\documentclass{llncs}
\usepackage{listings}
\usepackage{amsmath}
\usepackage{graphicx}
\begin{document}

%
\title{Optimised allgatherv, reduce\_scatter and allreduce communication in message-passing systems}

\author{Andreas Jocksch\inst{1} \and No\'{e} Ohana\inst{2} \and Emmanuel Lanti\inst{2} \and Vasileios Karakasis\inst{1} \and Laurent Villard\inst{2}}
\institute{
  CSCS, Swiss National Supercomputing Centre,\\ Via Trevano 131\\
  CH-6900 Lugano\\
  Switzerland
\and
  Ecole Polytechnique F\'ed\'erale de Lausanne (EPFL),\\ Swiss Plasma Center (SPC)\\
  CH-1015 Lausanne\\
  Switzerland
}

\maketitle

\begin{abstract}
Collective communications, namely the patterns {\tt allgatherv}, {\tt reduce\_scatter}, and {\tt allreduce} in message-passing systems are optimised based on measurements at the installation time of the library. The algorithms used are set up in an initialisation phase of the communication, similar to the method used in so-called persistent collective communication introduced in the literature. For {\tt allgatherv} and {\tt reduce\_scatter} the existing algorithms, recursive multiply/divide and cyclic shift (Bruck's algorithm) are applied with a flexible number of communication ports per node. The algorithms for equal message sizes are used with non-equal message sizes together with a heuristic for rank reordering. The two communication patterns are applied in a plasma physics application that uses a specialised matrix-vector multiplication. For the {\tt allreduce} pattern the cyclic shift algorithm is applied with a prefix operation. The data is gathered and scattered by the cores within the node and the communication algorithms are applied across the nodes. In general our routines outperform the non-persistent counterparts in established MPI libraries by up to one order of magnitude or show equal performance, with a few exceptions of number of nodes and message sizes.
\end{abstract}


\section{Introduction}
Collective communication is a component of the Message Passing Interface (MPI) library \cite{gropp_etal:1999}. While point-to-point communication provides basic functionality, collective communication can accommodate more complex algorithms inside the library. These algorithms can be very efficient with respect to the execution time \cite{thakur_etal:2005}. Persistent collective communication, which corresponds to communication patterns repeatedly called, plays an increasingly important role. It allows for even more sophisticated optimisations to be provided within the library, which are currently, with a setup of the algorithm in every communication call, inefficient due to the expensive initialisation phase. The implementation of persistent communication as non-blocking communication has been discussed in the literature \cite{hoefler_schneider:2012,mpi:2019,holmes_etal:2019}. In this contribution we introduce optimised collective communication for the patterns {\tt allgatherv}, {\tt reduce\_scatter}, and {\tt allreduce}, with the additional initialisation phase. Without restriction of the generality of our approach, our collectives are blocking. Overall we obtain speedups of up to one order of magnitude with respect to Cray MPI and a factor of five for MVAPICH.

More specifically, we generalise the recursive multiplication/division and cyclic shift algorithms for {\tt allreduce} and {\tt recursive\_scatter} to allow for different factors for different steps, as done in \cite{culler_etal:1993} for {\tt allreduce} only, using recursive exchange. In this way the underlying algorithms are adjusted for the particular network based on measurements. This procedure provides the majority of the performance improvement observed for the different cases.

Additionally, for {\tt allgatherv} and {\tt reduce\_scatter} with non-equal message sizes, we apply a heuristic for rank reordering in order to use the recursive multiplication/division and cyclic shift algorithms in an efficient way. At this point, we achieve speedups of $20\%$. The rank order is determined in the initialisation phase of the communication.

Furthermore, we generalise for {\tt allreduce} the prime factor decomposition for recursive multiplying \cite{culler_etal:1993} to a factorisation with multiple consecutive calls of Bruck's algorithm. The prime factors are combined using a greedy approach. For the case of the {\tt allreduce} operation, Bruck's algorithm is further optimised by storing the results of the prefix operation instead of the actual data. This allows shorter messages to be communicated.

Today, supercomputers are typically composed of many connected shared memory nodes, which provide fast communication between processor cores on the same node and slow communication between cores on different nodes. This property has been considered for optimising communication algorithms by several authors \cite{tipparaju_etal:2003,tu_etal:2008,li_etal:2014,venkatesh_etal:2014,li_etal:2018,jocksch_etal:2018,bayatpour_etal:2018}. Good speedups compared to standard implementations (MPICH, MVAPICH, OpenMPI) have been shown. However, only part of this work has been included in publicly available implementations. We exploit the shared memory on the node by gathering and scattering messages between cores on the node before and after sending them over the network, respectively, as done in the literature. In order to accommodate all optimisations efficiently, we generate a bytecode in the initialisation phase which is interpreted in the execution phase, as demonstrated in \cite{jocksch_etal:2018}.

Subsequently we present in Sec.~\ref{sec:background} the basic algorithms of our contribution and show in Sec.~\ref{sec:allgather} and \ref{sec:reduce_scatter_block} optimised routines for {\tt allgather} and \newline {\tt reduce\_scatter\_block}, respectively. The heuristic for non-equal message sizes is introduced in Sec.~\ref{sec:non_equal_message_sizes}. Furthermore in Sec.~\ref{sec:allreduce} an optimised routine for {\tt allreduce} is introduced. How the parameters of the algorithms are determined based on measurements at the library's installation time is discussed in Sec.~\ref{sec:parametrisation}. The implementation details are discussed in Sec.~\ref{sec:implementation}. Benchmarks made on a Dragonfly and on an Infiniband network are presented in Sec.~\ref{sec:benchmarks}. In Sec.~\ref{sec:fourier_filter} the routines {\tt allgatherv} and {\tt reduce\_scatter} are applied to matrix-vector multiplications of a Discrete Fourier Transform (DFT) Fourier filter of a plasma physics application, namely the ORB5 global electromagnetic Particle-In-Cell gyrokinetic turbulence code \cite{jolliet_etal:2007}. Related work is reviewed in Sec.~\ref{sec:related_work} and, finally, we draw our conclusions in Sec.~\ref{sec:conclusions}.

\section{Background}
\label{sec:background}
Several network topologies have emerged during the years; fat tree, hypercube, torus, and dragonfly are some examples. Beside their topology, networks are characterised by other properties like bandwidth and latency, which determine their performance. Simplified models like the logP model \cite{culler_etal:1993} are applied. The network properties lead to various different algorithms for collective communication, e.g. recursive multiplying, Bruck's algorithm \cite{bruck_etal:1997}, and the ring algorithm for {\tt allgather} operations, but also store and forward algorithms for personalised communication with small message sizes. The networks we are optimising for are the dragonfly network of a Cray XC40 KNL and an Infiniband network, although our algorithms, including the implementation, are also efficient on other network architectures.

For the optimisation of collective communication we consider mainly the literature about a fully connected network with a simple bandwidth-latency model for the communication cost. The network is assumed to have multiple ports for the communication. The ports are the connections of a node to the other nodes. All algorithms discussed are assumed to operate between nodes and only optionally between cores of the same node. Data exchange or data rearrangement within the node is assumed to have zero cost for our simple model.

\begin{figure}
\begin{center}
  \includegraphics[width=2cm]{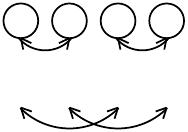}\hspace{1cm}
  \scalebox{-1}[1]{\includegraphics[width=2.2cm]{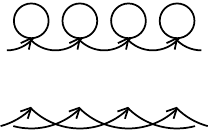}}
  \caption{Recursive multiplying/dividing (doubling/halfing; left) versus cyclic shift (also radix 2; right)}
  \label{fig:recursive}
\end{center}
\end{figure}

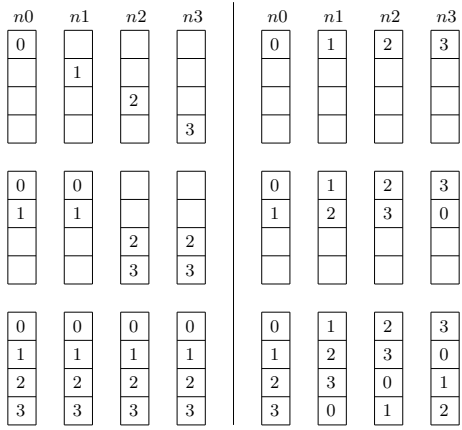
\begin{figure}
\begin{center}
  \scalebox{0.75}{
  \setlength{\unitlength}{1mm}
  \begin{picture}(80, 75)
    \multiput(0,0)(0,25){3}{
      \multiput(0,0)(45,0){2}{
        \multiput(0,0)(10,0){4}{\line(0,1){20}\line(1,0){5}}
        \multiput(5,20)(10,0){4}{\line(0,-1){20}\line(-1,0){5}}
        \multiput(0,5)(10,0){4}{\line(1,0){5}}
        \multiput(0,10)(10,0){4}{\line(1,0){5}}
        \multiput(0,15)(10,0){4}{\line(1,0){5}}
      }
      \put(1.5,16.5){0}
      \put(11.5,11.5){1}
      \put(21.5,6.5){2}
      \put(31.5,1.5){3}
    }
    \multiput(11.5,16.5)(0,25){2}{0}
    \multiput(1.5,11.5)(0,25){2}{1}
    \multiput(31.5,6.5)(0,25){2}{2}
    \multiput(21.5,1.5)(0,25){2}{3}
    \multiput(21.5,16.5)(10,0){2}{0}
    \multiput(21.5,11.5)(10,0){2}{1}
    \multiput(1.5,6.5)(10,0){2}{2}
    \multiput(1.5,1.5)(10,0){2}{3}
    \multiput(46.5,16.5)(0,25){3}{0}
    \multiput(56.5,16.5)(0,25){3}{1}
    \multiput(66.5,16.5)(0,25){3}{2}
    \multiput(76.5,16.5)(0,25){3}{3}
    \multiput(46.5,11.5)(0,25){2}{1}
    \multiput(56.5,11.5)(0,25){2}{2}
    \multiput(66.5,11.5)(0,25){2}{3}
    \multiput(76.5,11.5)(0,25){2}{0}
    \put(46.5,6.5){2}
    \put(56.5,6.5){3}
    \put(66.5,6.5){0}
    \put(76.5,6.5){1}
    \put(46.5,1.5){3}
    \put(56.5,1.5){0}
    \put(66.5,1.5){1}
    \put(76.5,1.5){2}
    \put(40,0){\line(0,1){75}}
    \multiput(1,71.5)(45,0){2}{$n0$}
    \multiput(11,71.5)(45,0){2}{$n1$}
    \multiput(21,71.5)(45,0){2}{$n2$}
    \multiput(31,71.5)(45,0){2}{$n3$}
  \end{picture}
  }
  \caption{Scheme of recursive multiplying (left) and cyclic shift (right), radix 2, initial data (top), after step 1 (middle), after step 2 (bottom), nodes $n0$-$n3$}
  \label{fig:scheme}
\end{center}
\end{figure}

The basic algorithms in this contribution are recursive multiplication/division and cyclic shift (Fig.~\ref{fig:recursive}).
Figure~\ref{fig:scheme} shows the data arrangement for the recursive multiplication and cyclic shift algorithms.
The top blocks show the buffers filled with the initial data. In the two execution steps for radix $2$ from top to bottom the buffers are filled further with the data communicated from the other nodes. The recursive multiplying algorithm has the advantage that with the last step the data is at the target and no local rearrangement on the node is necessary as for the cyclic shift algorithm. It is also an option to communicate the source data (top) to the destination (bottom) directly using 1 step with 3 so-called substeps. For recursive multiplication/division $s=\log_rp$ steps are required for $p$ nodes and a radix $r$, ($r^s=p$). Within a step $r-1$ messages need to be sent to different nodes which can be done in our nomenclature with $r-1$ ports.

\section{Adaptations of the algorithms}
We take the shared memory of the nodes into account and execute our algorithms according to the following steps (I) rearrangement of the data of all tasks on the node locally in a shared memory segment, (II) communication of the single node data to all nodes with our {\tt allgatherv}, {\tt reduce\_scatter}, or {\tt allreduce} algorithm, and (III) distribution of the data to all tasks on the node locally.

\subsection{Allgather}
\label{sec:allgather}
The allgather operation transmits from every participating rank, a piece of information, to every rank. Thus at the end of the operation every rank contains the same information which is the collected data of all ranks.

\begin{figure}
\begin{center}
  \includegraphics{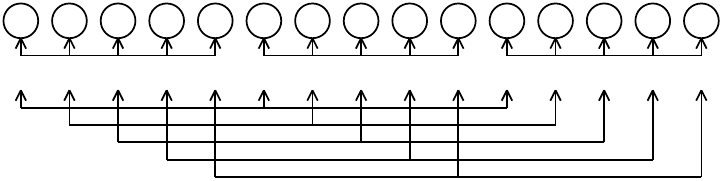}
  \caption{Recursive multiplying/dividing with factors 5 and 3}
  \label{fig:ring}
\end{center}
\end{figure}

Here we discuss equal message sizes (for non-equal ones see Sec.~\ref{sec:non_equal_message_sizes}). There are several options conceivable to perform the operation. In the literature, the most commonly used ones, which reduce the number of communication steps, are based on recursive multiplying (typically doubling) or cyclic shift (Bruck's algorithm \cite{bruck_etal:1997}), see Fig.~\ref{fig:recursive}. In difference to the naive algorithm which sends all information directly, these algorithms do not send the information directly from the source to the destination rank but apply forwarding. Thus the amount of data sent through the network remains unchanged, but the number of communication steps is reduced. In Fig.~\ref{fig:recursive} the algorithms are based on radix 2. At every step the information on every task is doubled, see Fig.~\ref{fig:scheme}: the data of both communication partners from before the step is on both partners after the step. We would like to emphasise that recursive multiplying and cyclic shift can be performed with radixes larger than two \cite{bruck_etal:1997,qian_afsahi:2007} or different factors for different steps. Figure~\ref{fig:ring} shows a communication done in two steps with factors 5 and 3.
We speak about factors $f_1\cdot f_2\cdot ...\cdot f_s=p$ since the formula with the radix $r^s=p$ is not valid in this case. The naive algorithm is equivalent to recursive multiplying or cyclic shift with the radix of number of nodes. Any combination of these algorithms seems to be possible, but we will not discuss the combination of cyclic shift and recursive multiplying further.

Within the steps the exchange of messages can be done in parallel with a number of ports equal to the factor minus one. Since the size of the messages grows from step to step we consider this flexibility as essential and use a different number of ports for the different steps of these two algorithms (see Sec.~\ref{sec:parametrisation}).

The algorithmic complexity for equal factors $f_i=r$ is
\begin{equation}
  T_{comm}=\alpha\log_rp+\beta((p-1)/(r-1)/p)n~,
\label{eq:comm_time_allgather}
\end{equation}
as shown in \cite{bruck_etal:1997}, where $T_{comm}$ is the time spent in communication, $n/p$ the number of bytes sent per node (assuming equal message sizes), $p$ the number of nodes participating, $r$ the radix of the algorithm, $\alpha$ the time required for a single step, and $\beta$ the time required for a single byte sent per node.

\subsection{Reduce\_scatter\_block}
\label{sec:reduce_scatter_block}
\begin{figure}
\begin{center}
  \scalebox{0.7}{
    \input{bruck_reverse.inc}
  }
  \caption{Cyclic shift for {\tt reduce\_scatter}, radix $r=2$, nodes $n0$-$n4$, numbers $0$-$4$ are messages to be reduced on the corresponding destination node, subscripts $0$-$4$ indicate the source node, $+$ is the reduction operation}
  \label{fig:bruck_reverse}
\end{center}
\end{figure}
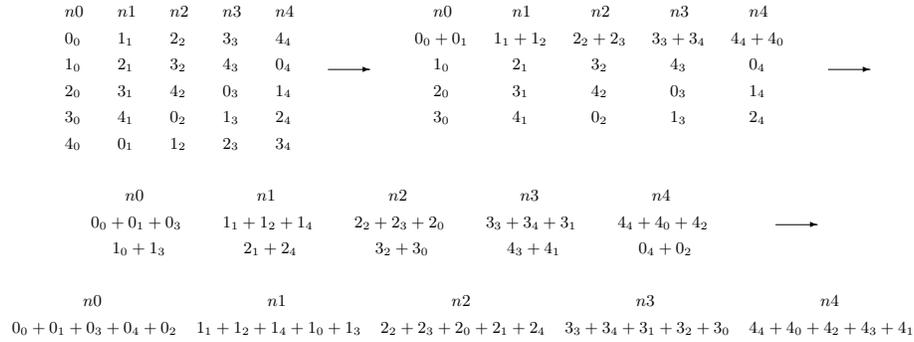

The optimisation of {\tt reduce\_scatter} operations has been the key aspect of several studies \cite{traff:2005,bernaschi_etal:2003}. For reduction operations one distinguishes between algorithms for commutative operations and non-commutative operations (see \cite{thakur_etal:2005} and references therein). In this contribution we consider commutative reduction operations only. The cyclic shift algorithm known from {\tt allgather} has also been applied in a similar way to {\tt reduce\_scatter\_block} \cite{bernaschi_etal:2003}. However, we believe that it should be formalised as the {\tt allgather} algorithm. The reduce\_scatter operation can be considered as the reversed {\tt allgatherv} operation in the same way as reduce is the reversed operation of broadcast. Therefore the same algorithms are applied in reversed order: Recursive division and cyclic shift. Figure~\ref{fig:bruck_reverse} shows an example of the reversed Bruck algorithm.
There is one major difference, however. While in the {\tt allgatherv} case buffers might be used for sending with multiple ports at the same time, this is only possible with an intra-node reduction for the {\tt reduce\_scatter} case. Thus the memory requirement is higher for {\tt reduce\_scatter}, since we assume that the receiving rank first gets the data in an empty buffer and second performs the arithmetic operation. Thus the memory requirements are increasing with an increasing number of ports. As the algorithms for {\tt allgatherv} and {\tt reduce\_scatter} differ in the direction of execution only, the algorithmic complexity is the same for both cases, except that the cost of reduction needs to be added for {\tt reduce\_scatter}. It is
\begin{equation}
  T_{comm}=\alpha\log_rp+\beta((p-1)/(r-1)/p)n+\gamma((p-1)/(r-1)/p)n~,
\label{eq:comm_time_reduce_scatter}
\end{equation}
where $\gamma$ is the computational cost per byte for the reduction.

\subsection{Non-equal message sizes}
\label{sec:non_equal_message_sizes}
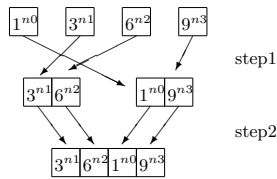
\begin{figure}
\begin{center}
  \scalebox{0.75}{
  \setlength{\unitlength}{1mm}
  \begin{picture}(70, 40)
    \put(10,30){\line(0,1){5}\line(1,0){5}}
    \put(15,35){\line(0,-1){5}\line(-1,0){5}}
    \put(10.5,31){$1^{n0}$}
    \put(20,30){\line(0,1){5}\line(1,0){5}}
    \put(25,35){\line(0,-1){5}\line(-1,0){5}}
    \put(20.5,31){$3^{n1}$}
    \put(30,30){\line(0,1){5}\line(1,0){5}}
    \put(35,35){\line(0,-1){5}\line(-1,0){5}}
    \put(30.5,31){$6^{n2}$}
    \put(40,30){\line(0,1){5}\line(1,0){5}}
    \put(45,35){\line(0,-1){5}\line(-1,0){5}}
    \put(40.5,31){$9^{n3}$}
    \put(12.5,17.5){\line(0,1){5}\line(1,0){5}}
    \put(17.5,22.5){\line(0,-1){5}\line(-1,0){5}}
    \put(13,18.5){$3^{n1}$}
    \put(17.5,17.5){\line(0,1){5}\line(1,0){5}}
    \put(22.5,22.5){\line(0,-1){5}\line(-1,0){5}}
    \put(18,18.5){$6^{n2}$}
    \put(32.5,17.5){\line(0,1){5}\line(1,0){5}}
    \put(37.5,22.5){\line(0,-1){5}\line(-1,0){5}}
    \put(33,18.5){$1^{n0}$}
    \put(37.5,17.5){\line(0,1){5}\line(1,0){5}}
    \put(42.5,22.5){\line(0,-1){5}\line(-1,0){5}}
    \put(38,18.5){$9^{n3}$}
    \put(17.5,5){\line(0,1){5}\line(1,0){5}}
    \put(22.5,10){\line(0,-1){5}\line(-1,0){5}}
    \put(18,6){$3^{n1}$}
    \put(22.5,5){\line(0,1){5}\line(1,0){5}}
    \put(27.5,10){\line(0,-1){5}\line(-1,0){5}}
    \put(23,6){$6^{n2}$}
    \put(27.5,5){\line(0,1){5}\line(1,0){5}}
    \put(32.5,10){\line(0,-1){5}\line(-1,0){5}}
    \put(28,6){$1^{n0}$}
    \put(32.5,5){\line(0,1){5}\line(1,0){5}}
    \put(37.5,10){\line(0,-1){5}\line(-1,0){5}}
    \put(33,6){$9^{n3}$}
    \put(12.5,30){\vector(2,-1){18}}
    \put(22.5,30){\vector(-1,-1){7}}
    \put(32.5,30){\vector(-2,-1){12}}
    \put(42.5,30){\vector(-1,-2){3}}
    \put(15,17.5){\vector(3,-4){5}}
    \put(20,17.5){\vector(3,-4){5}}
    \put(35,17.5){\vector(-3,-4){5}}
    \put(40,17.5){\vector(-3,-4){5}}
    \put(50,25){step1}
    \put(50,12){step2}
  \end{picture}
  }
  \caption{Pairing of messages/nodes, numbers $1$, $3$, $6$, and $9$ are sizes of the messages, superscripts $n0$-$n3$ are the nodes}
  \label{fig:pairing}
\end{center}
\end{figure}

\begin{figure}
\begin{center}
  \scalebox{0.75}{
  \vspace{7mm}
  \setlength{\unitlength}{1mm}
  \begin{picture}(7, 0)
    \put(0, 2){$n0$}
    \put(0, 9){$n1$}
    \put(0, 16){$n2$}
    \put(0, 23){$n3$}
    \put(9, 31){$\overbrace{\rule{12mm}{0mm}}^{T_{\text{step} 1}}$}
    \put(23.5, 31){$\overbrace{\rule{12mm}{0mm}}^{T_{\text{step} 2}}$}
    \put(85, 31){$\overbrace{\rule{12mm}{0mm}}^{T_{\text{step} 1}}$}
    \put(99.5, 31){$\overbrace{\rule{18mm}{0mm}}^{T_{\text{step} 2}}$}
  \end{picture}
  \includegraphics{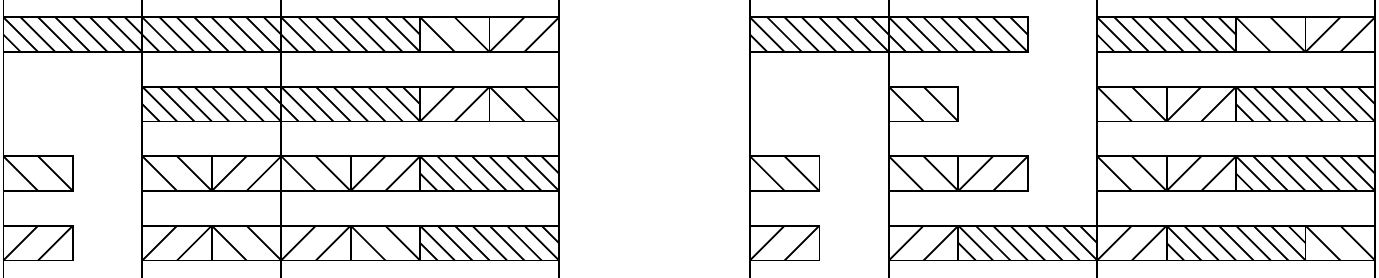}
  }
  \caption{Modelled execution for non-equal message sizes for recursive doubling (left) and cyclic shift (right) with radix 2; nodes $n0$-$n3$ with initial message sizes 1, 1, 0, 2; execution times of the two substeps $T_{step1}$ and $T_{step2}$ proportional to the message size; different hashes indicate different message tags; the longest message (horizontal extent) determines the communication time}
  \label{fig:gant}
\end{center}
\end{figure}
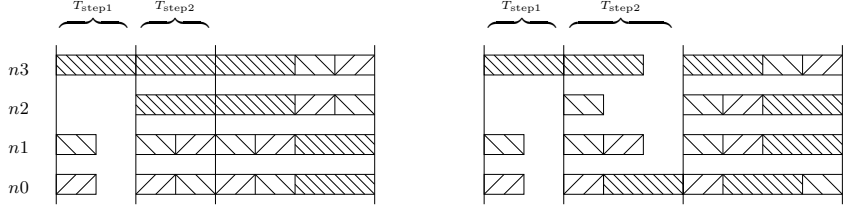

For the collective communications {\tt allgather} and {\tt reduce\_scatter\_block} with non-equal message sizes ({\tt allgatherv}, {\tt reduce\_scatter}), the principle that every rank performs the same number of operations with the same message sizes, which is due to symmetry, does not apply any more. This gives room for optimisations. However, in our approach we will leave the basic algorithms unmodified. We exploit the option of rank reordering for the algorithm (not for the network). Our heuristic for non-equal message sizes is to pair small messages with large messages in the different communication steps (Fig.~\ref{fig:pairing}).
The different ranks are grouped in a tree like order. For every communication step for an odd number of messages the largest message is taken out and remains. For the rest of the messages, as for an even number of messages, the smallest one will be paired with the largest one, the second smallest one with the second largest one, and so on. The two messages within one pair are sorted. The sums of the message sizes of the pairs become the message sizes of the next step. For example in Fig.~\ref{fig:pairing} the nodes will be ordered $n1$, $n2$, $n0$, $n3$. While for equal message size recursive multiplying and cyclic shifting requires the same execution time, for non-equal message sizes it does not. The example in Fig.~\ref{fig:gant} shows both algorithms applied to reordered messages of size 1, 1, 0, 2 at the start of the communication with an communication time $T_{comm}=4$ for recursive multiplying (left) and $T_{comm}=5$ the cyclic shift (right), assuming zero latency and a bandwidth of one. In this case, but not in general, all arrangements other than the one presented for the recursive multiplying algorithm give $T_{comm}=5$, and for the cyclic shift all arrangements give $T_{comm}=5$.

If zero message sizes occur, the algorithms can be interpreted in an alternative way. An example is the ordered messages of size $0$, $1$, $0$, $1$, $0$, $1$, $0$, $1$. The algorithm with radix 2 corresponds to four scatter operations between pairs of two nodes followed by two parallel allgather operations with four members each.

In the current implementation the rank reordering procedure is executed redundantly on all nodes for each pairing step using the quicksort algorithm. Thus the algorithmic complexity of the initialisation phase is
\begin{equation}
  T_{reorder}=\delta p(\log_2p)^2~,
\label{eq:complexity_reorder}
\end{equation}
where $\delta$ is the time required for a basic operation -- compare and swap if necessary -- of the sorting algorithm and $p$ the number of nodes. Other solutions for the sorting as a distributed sorting can be used \cite{jeon_kim:2003}, which become relevant for many MPI tasks \cite{balaji_etal:2009}, in order to reduce the algorithmic complexity of the initialisation. Otherwise the initialisation would dominate from a very large number of tasks and above (see Eqns.~(\ref{eq:comm_time_allgather}), (\ref{eq:comm_time_reduce_scatter}) and (\ref{eq:complexity_reorder})).

\subsection{Allreduce}
\label{sec:allreduce}
We follow the literature and base our allreduce collective operation for small messages on the same algorithm as {\tt allgatherv}. A naive implementation would use the {\tt allgather} algorithm without any modification as illustrated in Figure~\ref{fig:allreduce_bruck} (left) for cyclic shift.
\begin{figure}
\begin{center}
  \scalebox{0.75}{
  \setlength{\unitlength}{1mm}
  \begin{picture}(60,60)
     \multiput(6.5,1.5)(5,5){11}{$A$}
     \multiput(6.5,6.5)(5,5){10}{$9$}
     \multiput(11.5,1.5)(5,5){10}{$0$}
     \multiput(6.5,11.5)(5,5){9}{$8$}
     \multiput(16.5,1.5)(5,5){9}{$1$}
     \multiput(6.5,16.5)(5,5){8}{$7$}
     \multiput(21.5,1.5)(5,5){8}{$2$}
     \multiput(6.5,21.5)(5,5){7}{$6$}
     \multiput(26.5,1.5)(5,5){7}{$3$}
     \multiput(6.5,26.5)(5,5){6}{$5$}
     \multiput(31.5,1.5)(5,5){6}{$4$}
     \multiput(6.5,31.5)(5,5){5}{$4$}
     \multiput(36.5,1.5)(5,5){5}{$5$}
     \multiput(6.5,36.5)(5,5){4}{$3$}
     \multiput(41.5,1.5)(5,5){4}{$6$}
     \multiput(6.5,41.5)(5,5){3}{$2$}
     \multiput(46.5,1.5)(5,5){3}{$7$}
     \multiput(6.5,46.5)(5,5){2}{$1$}
     \multiput(51.5,1.5)(5,5){2}{$8$}
     \multiput(6.5,51.5)(5,5){1}{$0$}
     \multiput(56.5,1.5)(5,5){1}{$9$}
     \put(5.5,56.5){$n0$}
     \put(10.5,56.5){$n1$}
     \put(15.5,56.5){$n2$}
     \put(20.5,56.5){$n3$}
     \put(25.5,56.5){$n4$}
     \put(30.5,56.5){$n5$}
     \put(35.5,56.5){$n6$}
     \put(40.5,56.5){$n7$}
     \put(45.5,56.5){$n8$}
     \put(50.5,56.5){$n9$}
     \put(55.5,56.5){$nA$}
     \put(5,0){\line(1,0){55}}
     \put(5,15){\line(1,0){55}}
     \put(5,35){\line(1,0){55}}
     \put(5,45){\line(1,0){55}}
     \put(5,50){\line(1,0){55}}
     \put(0.5,0.5){\line(1,1){4}}
     \put(4.5,0.5){\line(-1,1){4}}
     \put(0.5,15.5){\line(1,1){4}}
     \put(4.5,15.5){\line(-1,1){4}}
     \put(0.5,35.5){\line(1,1){4}}
     \put(4.5,35.5){\line(-1,1){4}}
     \put(0.5,40.5){\line(1,1){4}}
     \put(4.5,40.5){\line(-1,1){4}}
     \put(0.5,45.5){\line(1,1){4}}
     \put(4.5,45.5){\line(-1,1){4}}
     \put(0.5,50.5){\line(1,1){4}}
     \put(4.5,50.5){\line(-1,1){4}}
  \end{picture}\hspace{1cm}
  \begin{picture}(70,60)
     \put(0,51.5){$\sum_{i=0}^0i$}
     \put(0,46.5){$\sum_{i=0}^1i$}
     \put(0,41.5){$\sum_{i=0}^2i$}
     \put(0,36.5){$\sum_{i=0}^3i$}
     \put(2,31.5){$-$}
     \put(2,26.5){$-$}
     \put(2,21.5){$-$}
     \put(0,16.5){$\sum_{i=0}^7i$}
     \put(2,11.5){$-$}
     \put(2,6.5){$-$}
     \put(0,1.5){$\sum_{i=0}^Ai$}
     \put(19,51.5){$\sum_{i=1}^1i$}
     \put(19,46.5){$\sum_{i=1}^2i$}
     \put(19,41.5){$\sum_{i=1}^3i$}
     \put(19,36.5){$\sum_{i=1}^4i$}
     \put(21,31.5){$-$}
     \put(21,26.5){$-$}
     \put(21,21.5){$-$}
     \put(19,16.5){$\sum_{i=1}^8i$}
     \put(21,11.5){$-$}
     \put(21,6.5){$-$}
     \put(14,1.5){$\sum_{i=1}^Ai+\sum_{i=0}^0i$}
     \put(55,51.5){$\sum_{i=A}^Ai$}
     \put(50,46.5){$\sum_{i=A}^Ai+\sum_{i=0}^0i$}
     \put(50,41.5){$\sum_{i=A}^Ai+\sum_{i=0}^1i$}
     \put(50,36.5){$\sum_{i=A}^Ai+\sum_{i=0}^2i$}
     \put(58,31.5){$-$}
     \put(58,26.5){$-$}
     \put(58,21.5){$-$}
     \put(50,16.5){$\sum_{i=A}^Ai+\sum_{i=0}^6i$}
     \put(58,11.5){$-$}
     \put(58,6.5){$-$}
     \put(50,1.5){$\sum_{i=A}^Ai+\sum_{i=0}^9i$}
     \put(2,56.5){$n0$}
     \put(21.5,56.5){$n1$}
     \put(40,56.5){$...$}
     \multiput(40,1.5)(0,5){11}{$...$}
     \put(58,56.5){$nA$}
     \put(0,0){\line(1,0){70}}
     \put(0,15){\line(1,0){70}}
     \put(0,35){\line(1,0){70}}
     \put(0,45){\line(1,0){70}}
     \put(0,50){\line(1,0){70}}
  \end{picture}
  }
  \caption{Cyclic shift algorithm adapted from {\tt allgather}, nodes $n0$-$nA$ with messages $0$-$A$ (hexadecimal notation), radix 2, horizontal lines \rule{7mm}{0.15mm} indicate the end of every step, $X$ are the lines required for {\tt allreduce} (left), for sum reduction inclusive scans from top to bottom which are actually stored and communicated (right)}
  \label{fig:allreduce_bruck}
\end{center}
\end{figure}
We assume that the reduction operation is a sum. Here, we modify the scheme and do not store the values at the lines but column-wise the partial sum (inclusive scan) from the top to the bottom (Fig.~\ref{fig:allreduce_bruck} right). While the $l$'th line shifted by $k$ columns to the left is the $l+k$'th line in the original scheme (Fig.~\ref{fig:allreduce_bruck} left), in our modified version, for a block of lines from $1$st to $n$'th shifted by $k$ columns to the left and $k$ lines to the bottom, the prefix sum for the shifted lines is computed by adding the prefix sum of line $k-1$ (Fig.~\ref{fig:allreduce_bruck} right).

This idea allows for less lines to be communicated, since for computing the final result on the bottom line, for $r=2$ only the lines which are marked with an $X$ on the left are required, the rest of the lines are not needed and are displayed for the illustration of the algorithm only. In case of complete steps, e.g. for a radix of $r=2$ and $2^n$ nodes the algorithmic complexity becomes equivalent to the one of the binary exchange algorithm. In the more general case, if always $(f_1-1)\cdot (f_2-1)\cdot ...\cdot (f_i-1)$ lines according to the non-optimised algorithm are communicated including the last step, only one line per substep needs to be communicated. This case corresponds to the approach of \cite{ruefenacht_etal:2017} where the node count is decomposed in a product of prime numbers. The message exchange is done for each single factor separately with a reduction followed. In \cite{ruefenacht_etal:2017} this special case was applied to the recursive exchange algorithm.

For non $2^n$ nodes but a radix $r=2$ more lines need to be communicated in every step than the last line. This is due to the incomplete last step of the cyclic shift. The additional lines double at most the data volume. The generalisation to non-equal arbitrary factors is straightforward. It might be efficient to set $f_i>2$ (see Sec.~\ref{sec:parametrisation}).

As discussed already, the {\tt allreduce} communication can be executed independently for every factor of the node count. Therefore the node count is decomposed in prime factors. If the prime factors are smaller than a target factor $f_i$ (e.g. $f_i=13$) they are combined according to a greedy approach. For prime factors much larger than the target factor $f_i$ e.g. $f_i=167$ we apply cyclic shift operations with multiple steps for every prime factor i.e. two factors 13 for 167.

For long messages we use Rabenseifner's algorithm \cite{thakur_etal:2005} and perform a \newline {\tt reduce\_scatter} followed by an {\tt allgatherv}. These two routines were discussed in the previous sections. With the cyclic shift algorithm for these routines, we are not bound to any particular node count, such as the $2^n$ used in the literature. In this case the algorithmic complexity is the sum of the complexity of {\tt reduce\_scatter} and {\tt allgatherv}.

Alternative algorithms have been described in the literature, e.g. the binary blocks algorithm \cite{thakur_etal:2005}. This algorithm considers the node count as sum of $2^{n_i}$ parts which are executed independently and are reduced with each other at the end. For example, it requires 4 steps for 7 nodes. For very short messages our algorithm corresponds to {\tt allgather} and it outperforms the binary block algorithm since it requires less steps (3 steps for 7 nodes). Comparisons of all message sizes and possible combinations of the algorithms are still under investigation.

\section{Parametrisation}
\label{sec:parametrisation}
The simple bandwidth-latency network model does not give any indication which factors $f_i$ (or radix $r$) to choose for bandwidth dominated communication. Furthermore it is not always clear how many ports per node are available. In order to choose the optimal parameters we apply a tuning approach. At the installation phase of the library, measurements of communication times are done for different message sizes. Based on that, the factors $f_i$ are chosen. For all possible combinations of factors the communication time is estimated from interpolations of the measurements performed during installation. The algorithmic complexity of this try-all method is
\begin{equation}
  N_{op}\le f_{i,max}^{\log_2p},\ \ p=2^n,\ \ n\in \mathbf{N}
\end{equation}

The total number of messages between nodes is in any case smaller with our shared memory approach than for a naive implementation, since messages are merged. This is advantageous with respect to network congestion. The option of splitting the messages between nodes and using multiple senders and receivers for their transmission is not exploited here. However, the load on the network might still affect the communication. That is why, the measurement runs are done with different loads on the network in the background using the GPCNeT benchmark \cite{gpcnet:2019}. The parameters of the algorithms can be adapted to the network load.

Our experiments show that it is efficient to apply high and low radixes for short messages and long messages, respectively. This is supported by the findings in \cite{parsons:2015}, where a saturation effect for long messages is described. The backload of the network boosts this effect.

The $r-1$ ports can be physical ports in the sense of multiple cores performing communication or logical ports if one core performs multiple non-blocking point-to-point communications. In this contribution we restrict ourselves to physical ports (see also \cite{jocksch_etal:2018}). If the factors $f_i$ allow, the recursive multiply/divide is applied, otherwise the cyclic shift, since the former seems to be advantageous for non-equal message sizes (Sec.~\ref{sec:non_equal_message_sizes}). We apply one exception to the tuning approach: the target factor $f_i$ is fixed to the number of cores per node plus one for {\tt allreduce} with small message sizes.

\section{Implementation details}
\label{sec:implementation}
The separation of the initialisation phase of the algorithms from the actual communication is beneficial since a significant amount of computation has to be done in order to determine the parameters, algorithms, single step message sizes and communicating ranks. The execution time of the initialisation is approximately independent from the message size, and thus not negligible especially for short messages (see Sec.~\ref{sec:benchmarks}). The cost of initialisation is amortised by repeated calls of the execution routines which are highly optimised. Therefore we have chosen to encode the whole algorithm in a special bytecode in the initialisation phase, without any ifs/jumps \cite{jocksch_etal:2018}. In the execution phase this bytecode is interpreted. We have many algorithmic choices in the code generation phase, without disadvantage in the execution phase.

Our collective communication routines are based on the MPI point-to-point communication routines MPI\_Irecv, MPI\_Isend, MPI\_Waitall and MPI\_Sendrecv. The latter one is used for all {\tt reduce\_scatter} implementations only.

Since the algorithms are purely deterministic, numerical results of the reductions are bit-reproducible.

Current limitations of the implementation are the following. The same number of cores must be involved in the collective communication on all nodes. It is not allowed to let the last node partially idle if a certain overall number of cores is required, which is a realistic use pattern. In the automatic determination of the algorithmic parameters, no tradeoff between performance and memory usage is implemented. Inter-node communicators \cite{kang_etal:2018} are not implemented yet. Non-contiguous data types are not supported.

It is not necessary to rewrite our routines for equal message sizes, {\tt allgather} and {\tt reduce\_scatter\_block}, since they would not perform better than the versions for non-equal message sizes, but wrappers might be convenient for this feature. Furthermore the operations {\tt bcast} and {\tt reduce} are covered by setting up {\tt allgatherv} and {\tt reduce\_scatter}, respectively, with all message sizes equal to zero except of one. Then our algorithms simplify to a tree algorithm without explicit implementation. In difference to the interface planned in MPI 4.0, which foresees non-blocking persistent collective communication only, our collective communication is blocking. For the current blocking implementation we see various applications already besides the one of plasma physics shown.

The source code of our collective communication implementation will be made publicly available on github if the contribution will be accepted.

\section{Benchmarks}
\label{sec:benchmarks}
Benchmarks are made on an empty Cray XC40 KNL cluster comprising 64-core Intel(R) Xeon Phi(TM) CPU 7230 processors running at 1.30GHz. The processors are configured in flat memory mode with quadrant clustering using MCDRAM. The network topology is Dragonfly with Aries routing. Furthermore we utilise an empty Infiniband cluster using 17 nodes with two 12 core Haswell(TM) E5-2650v.3 2.66 GHz CPUs.

\begin{figure}
\begin{center}
  \includegraphics[width=5.5cm]{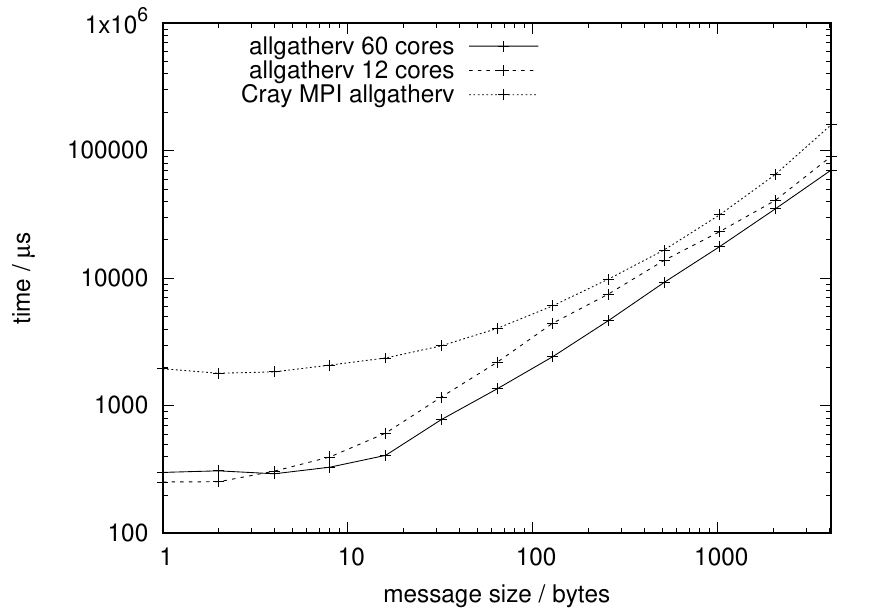}\hspace{0.25cm}
  \includegraphics[width=5.5cm]{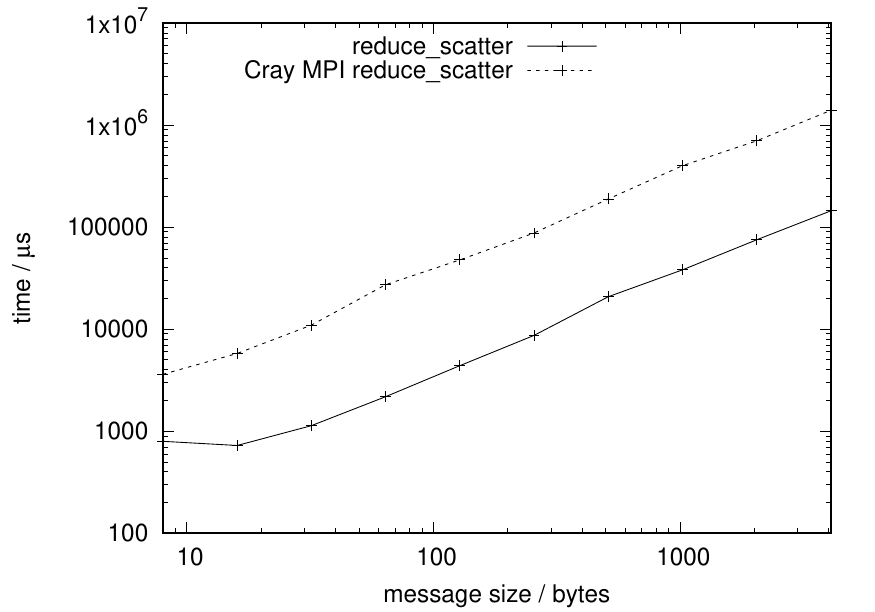}\\
  \includegraphics[width=5.5cm]{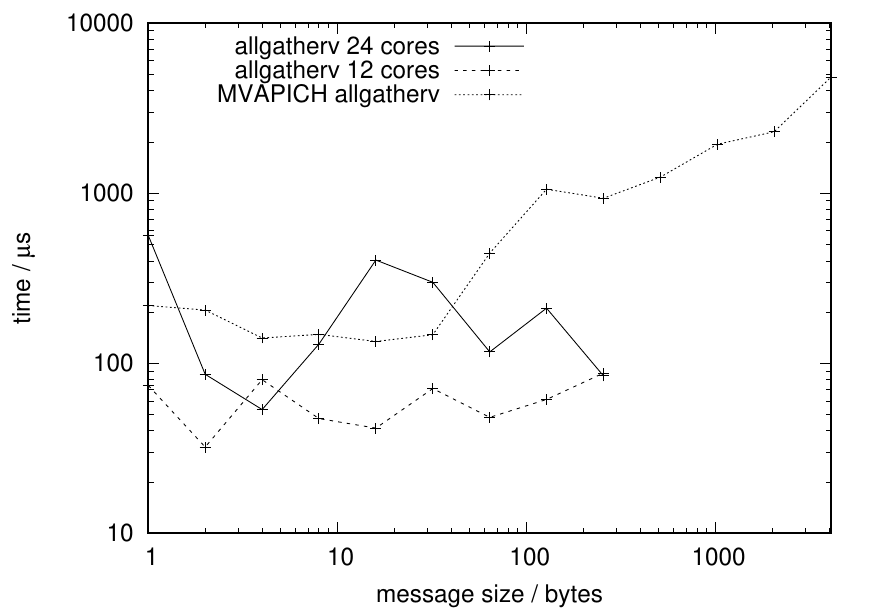}\hspace{0.25cm}
  \includegraphics[width=5.5cm]{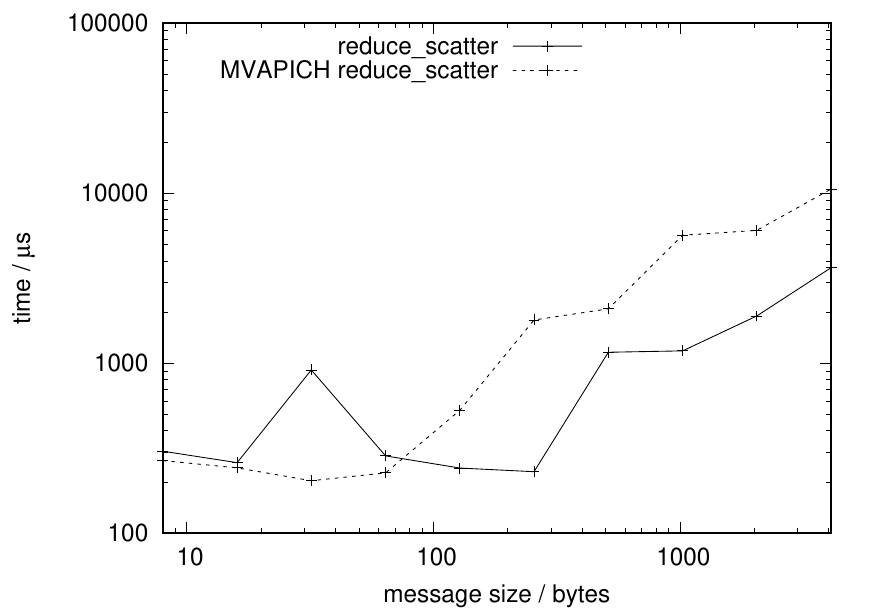}
  \caption{Allgatherv (left) and Reduce\_scatter (right) on 160 nodes with 9600 tasks Cray (top) and on 17 nodes with 408 tasks Infiniband (bottom)}
  \label{fig:allgatherv_reduce_scatter}
\end{center}
\end{figure}

\begin{figure}
\begin{center}
  \includegraphics[width=5.5cm]{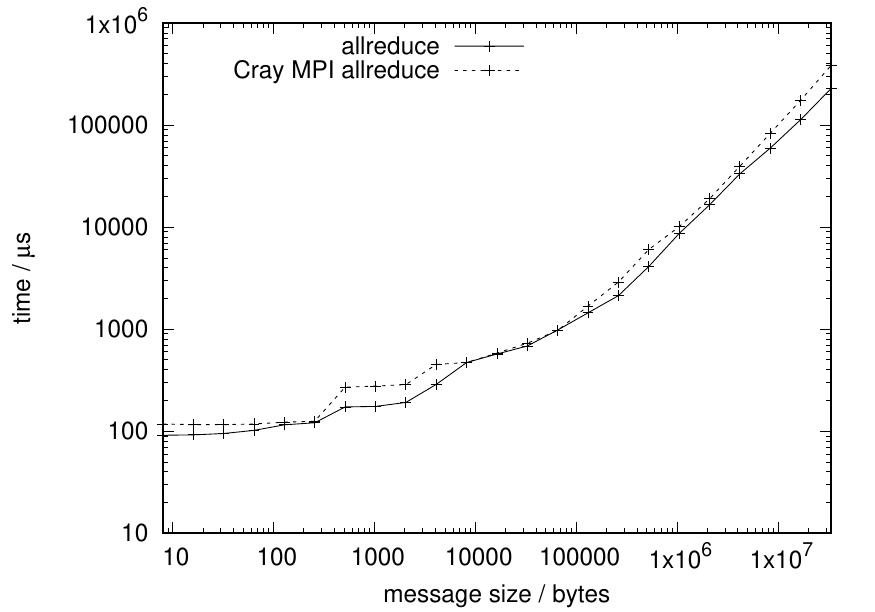}\hspace{0.25cm}
  \includegraphics[width=5.5cm]{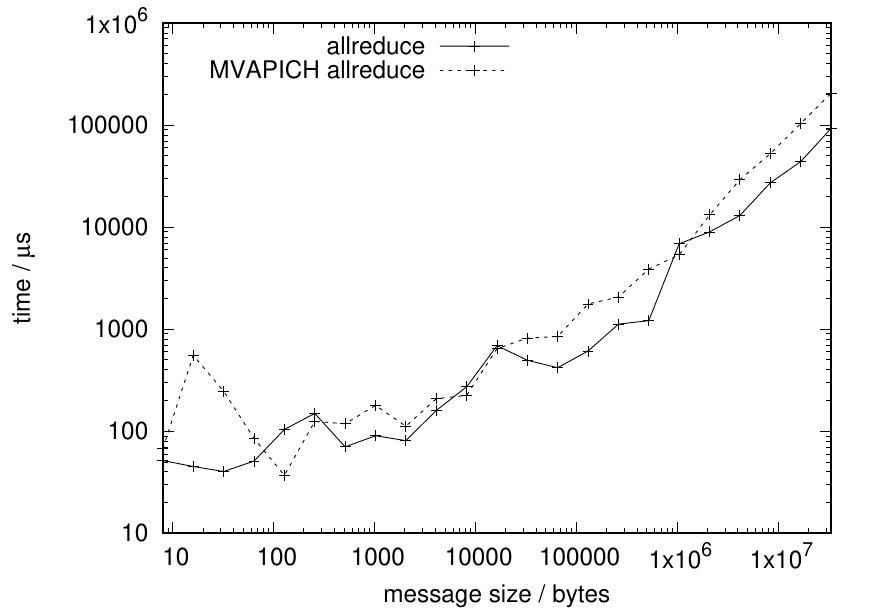}
  \caption{Allreduce on 160 nodes with 9600 tasks Cray (left) and on 17 nodes with 408 tasks Infiniband (right)}
  \label{fig:allreduce}
\end{center}
\end{figure}

\begin{figure}
\begin{center}
  \includegraphics[width=5.5cm]{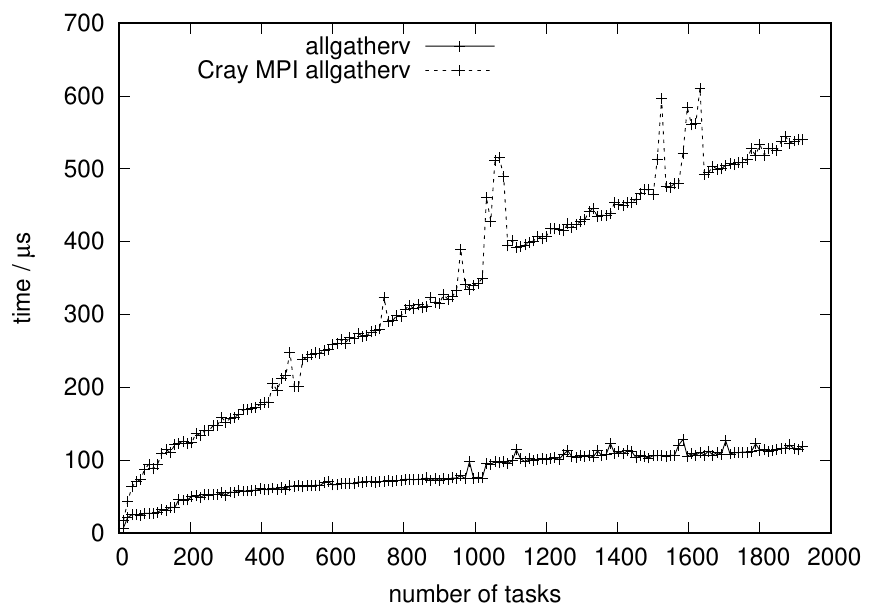}\hspace{0.25cm}
  \includegraphics[width=5.5cm]{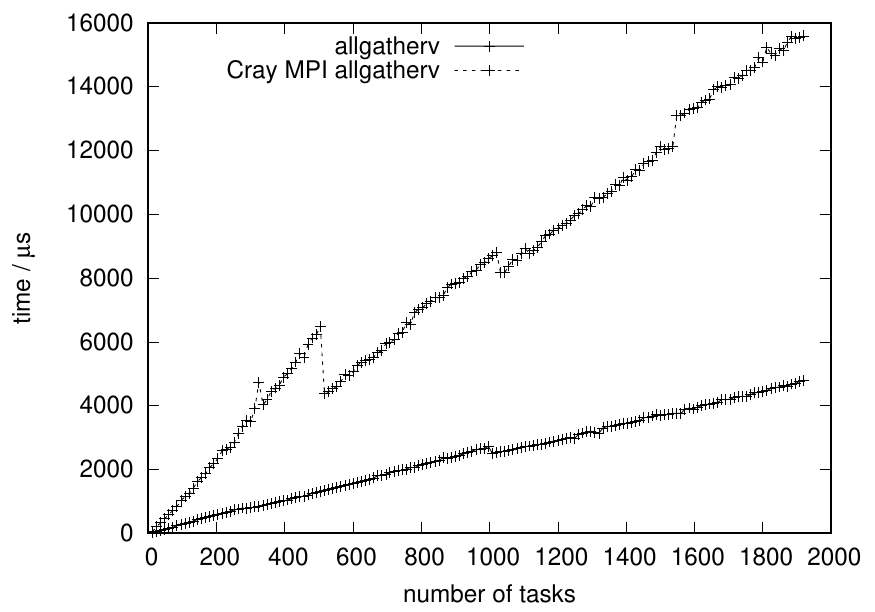}
  \caption{Allgatherv with 8 bytes (left) and 4096 bytes (right)}
  \label{fig:allgatherv_b}
\end{center}
\end{figure}

\begin{figure}
\begin{center}
  \includegraphics[width=5.5cm]{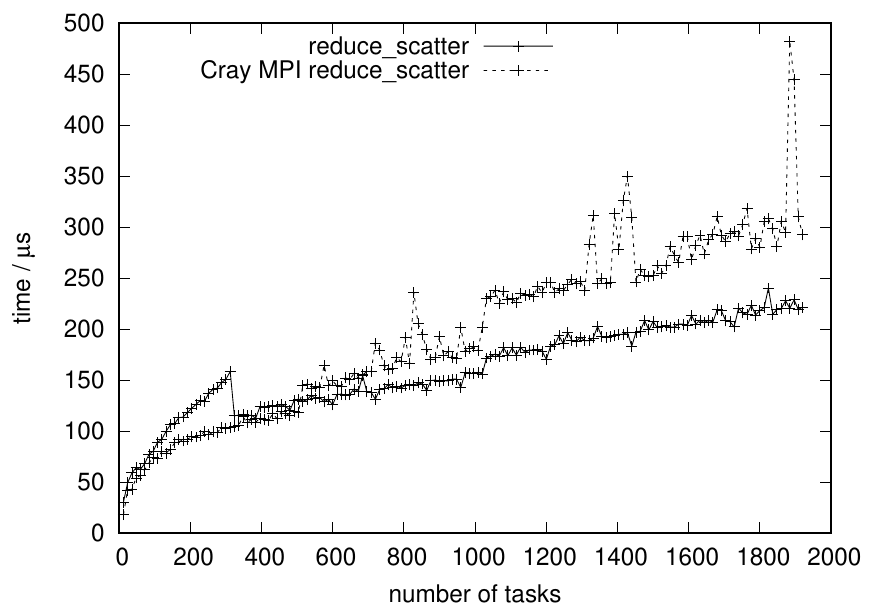}\hspace{0.25cm}
  \includegraphics[width=5.5cm]{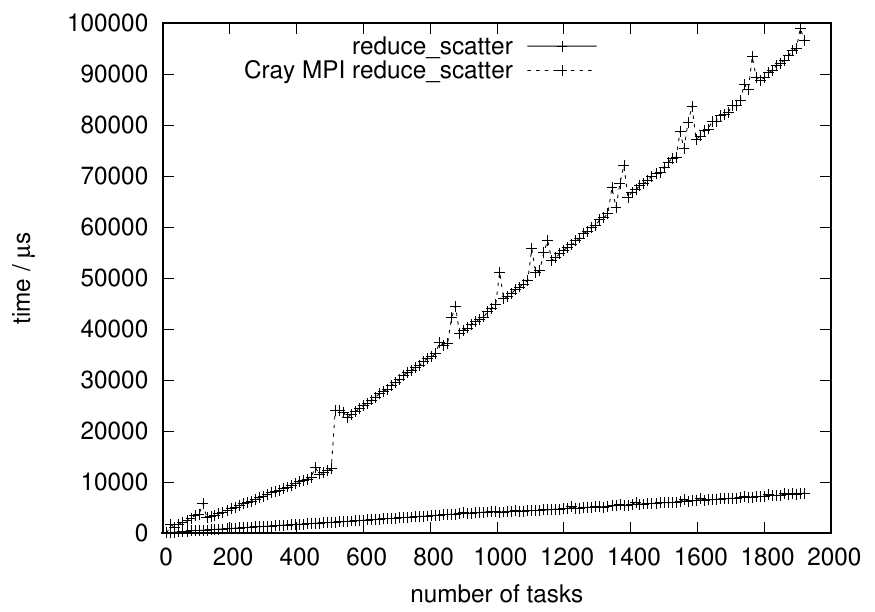}
  \caption{Reduce\_scatter with 8 bytes (left) and 4096 bytes (right)}
  \label{fig:reduce_scatter_b}
\end{center}
\end{figure}

\begin{figure}
\begin{center}
  \includegraphics[width=5.5cm]{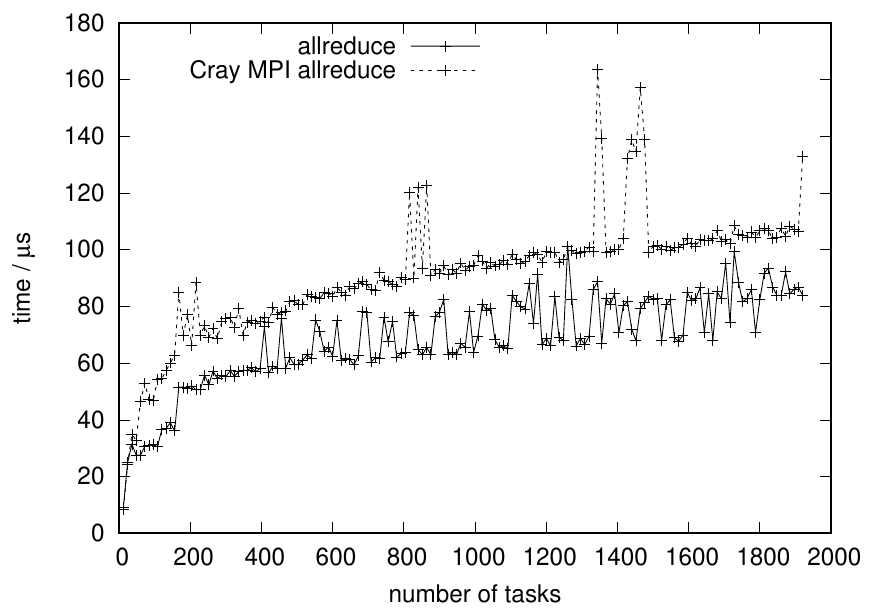}\hspace{0.25cm}
  \includegraphics[width=5.5cm]{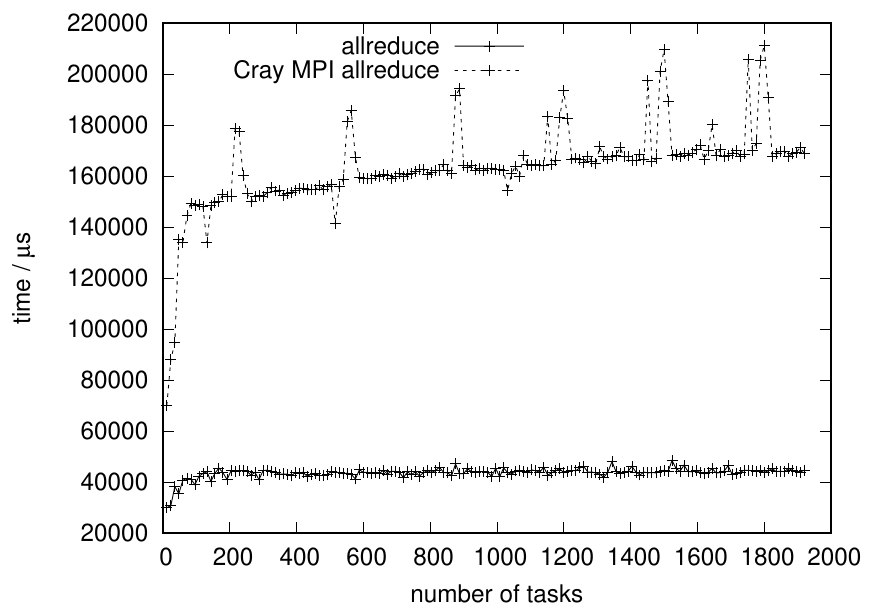}
  \caption{Allreduce with 8 bytes (left) and 33554432 bytes (right)}
  \label{fig:allreduce_b}
\end{center}
\end{figure}

We follow the OSU microbenchmarks \cite{bureddy_etal:2012}, which were adapted for our communication routines. Figure~\ref{fig:allgatherv_reduce_scatter} (left) shows the communication time in relation to the message size for our persistent {\tt allgatherv} routine and for the non-persistent MPI {\tt allgatherv} routine for 9600 cores on 160 nodes.
For our routine, the Cray nodes were used in two different ways: either groups of 12 cores each, forming 5 virtual nodes per KNL or one group of 60 cores per KNL. The 12 cores per virtual node were chosen in order to mimic systems with 12 cores per node. The message sizes refer to the message of the send buffer before communication. Our routine is faster than the one of Cray MPI (cray-mpich/7.7.10) on the Cray network especially for small message sizes. For very large message sizes (not shown) the performance becomes equal between the two implementations. For the Infiniband (Inf) network our routine mostly outperforms MVAPICH 2.2, for both one group of 24 cores per node and two groups of 12 cores per node. Figure~\ref{fig:allgatherv_reduce_scatter} (right) shows the same properties for {\tt reduce\_scatter} while from here on we restrict ourselves to virtual nodes (per KNL / dual socket Haswell) with 12 cores each for all subsequent benchmarks. Contrary to the definition of the OSU microbenchmarks, the message size refers to the message in the receive buffer after communication. The speedup of our routine compared to Cray MPI (Cray network) and to MVAPICH (Inf network) is significant. We believe that, besides our algorithmic improvements implemented for the communication, also the local reduction on the node is done more efficiently in our routines. Figure~\ref{fig:allreduce} shows the results for {\tt allreduce}.
For small and large message sizes our routines outperform the Cray MPI, but for medium message sizes (about 10kB-100kB) the routines show approximately equal performance. Our routine is mostly faster than MVAPICH. For {\tt allgatherv} on the Cray network for the smallest message size the initialisation is 5700 times more expensive than the single execution of the algorithm, for the longest message it is a factor of 56. Figures~\ref{fig:allgatherv_b},\ref{fig:reduce_scatter_b}, and \ref{fig:allreduce_b} show the performance of the different routines for a varying number of nodes on the Cray network with a fixed message size. Our results are comparable with the speedups of End et al. \cite{end_etal:2016} who have implemented a k-port allreduce and have made a comparison with OpenMPI 1.6.5.
Our routines outperform Cray MPI in all cases except {\tt reduce\_scatter} at small message sizes for a few number of nodes. The peaks in all graphs especially in the ones for small messages sizes are intermittent slowdowns of the system and not caused by the algorithms.

\section{Fourier filter}
\label{sec:fourier_filter}
\begin{figure}
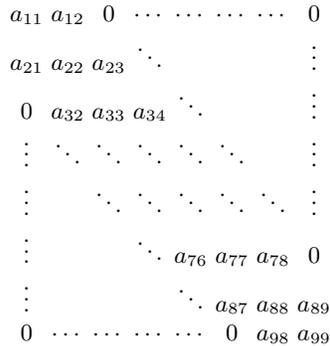

\begin{equation*}
\begin{matrix}
a_{11}  & a_{12}  & 0 & \cdots & \cdots & \cdots & \cdots & 0 \\
a_{21}  & a_{22}  & a_{23}  & \ddots & && & \vdots \\
0 & a_{32}  & a_{33} & a_{34}  & \ddots & &  & \vdots \\
\vdots & \ddots & \ddots & \ddots & \ddots & \ddots &  & \vdots \\
\vdots & & \ddots & \ddots & \ddots & \ddots & \ddots& \vdots\\
\vdots  &  & & \ddots & a_{76}  & a_{77}  &  a_{78}  & 0\\
\vdots  &  & & & \ddots & a_{87}  & a_{88}  &  a_{89}\\
0 & \cdots &  \cdots & \cdots & \cdots & 0 & a_{98} & a_{99}  \\
\end{matrix}
\end{equation*}
\caption{Fourier modes retained in the $r=const.$ plane}
\label{fig:fourier_mode_matrix}
\end{figure}

The optimised allgather and reduce\_scatter routines are applied to a Fourier filter which is part of the plasma physics application ORB5 \cite{jolliet_etal:2007}. Its task is to transform data on a regular 3D mesh which is periodic in two directions from real space to spectral space in these two periodic directions and to select a fraction of modes to be processed further. The reverse spectral space to real space transformation is also part of the procedure. The data arrangement of the code is the following. The application uses a toroidal computational domain, for parallelisation a 1D domain decomposition in toroidal direction, and an additional domain cloning technique. The filter reduces the number of Fourier modes to a band in poloidal-toroidal mode numbers. For general configurations the number of Fourier modes processed further in the field solver is not a multiple of the number of nodes allocated, the messages have non-equal size. It might even happen that part of the nodes are idling during the field solve procedure and will either receive or send messages only. Figure~\ref{fig:fourier_mode_matrix} illustrates the Fourier modes retained or set to zero for a certain radial coordinate.
The filter varies in radial direction. Two options are implemented in the code, a solution with Fast Fourier Transforms (FFTs) and one with a DFT matrix, other ones are possible.

In the DFT matrix approach the real space vector $\mathbf{r}$ is transformed to the spectral space vector $\mathbf{s}$ by multiplying with the matrix $\mathbf{F}$.
\begin{equation}
\mathbf{s}=\mathbf{F}\mathbf{r}
\end{equation}
This operation with $N^2$ complexity is efficient in our case since the transformation matrix $\mathbf{F}$ is very sparse. Here the start is a FFT in poloidal direction followed by the matrix-vector multiplication. Thus the matrix
\begin{equation}
\mathbf{F}=\begin{bmatrix}&\scalebox{1.5}{0}\\\hline\omega_N^{l\cdot 0}&\omega_N^{l\cdot 1} &\hdots &\omega_N^{l\cdot(N-1)}\\\vdots &\vdots &&\vdots\\\omega_N^{m\cdot 0}&\omega_N^{m\cdot 1} &\hdots &\omega_N^{m\cdot(N-1)}\\\hline &\scalebox{1.5}{0}\end{bmatrix}~,
~\omega_N=e^{-i2\pi/N}
\end{equation}
transforms a single line in toroidal direction. Only the values necessary to be computed are communicated. The computation and communication from $\mathbf{r}$, which is distributed over the nodes, to $\mathbf{s}$ is done such that $\mathbf{s}$ is distributed as equal as possible over the nodes. For the backward transformation the reverse operations apply.

\begin{figure}
\begin{center}
  \includegraphics[width=5.5cm]{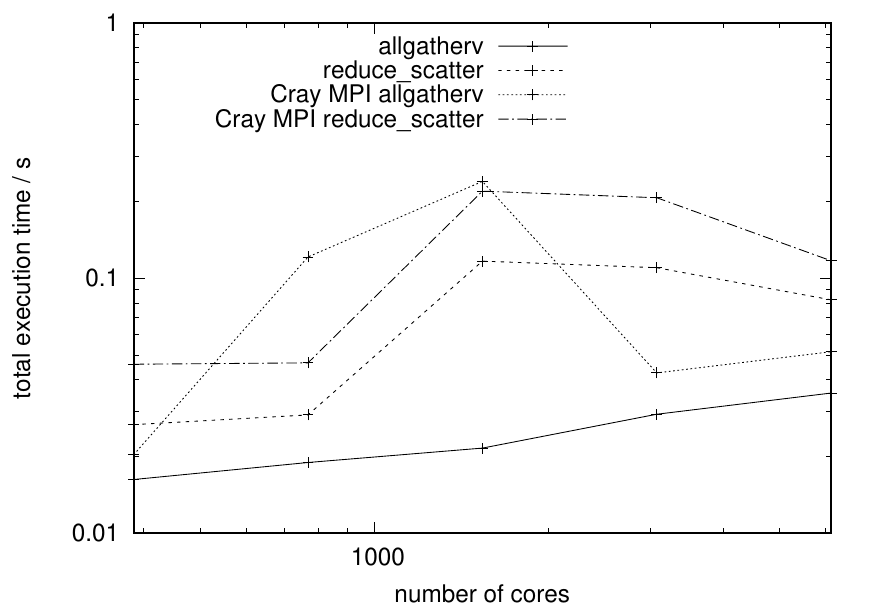}\hspace{0.25cm}
  \includegraphics[width=5.5cm]{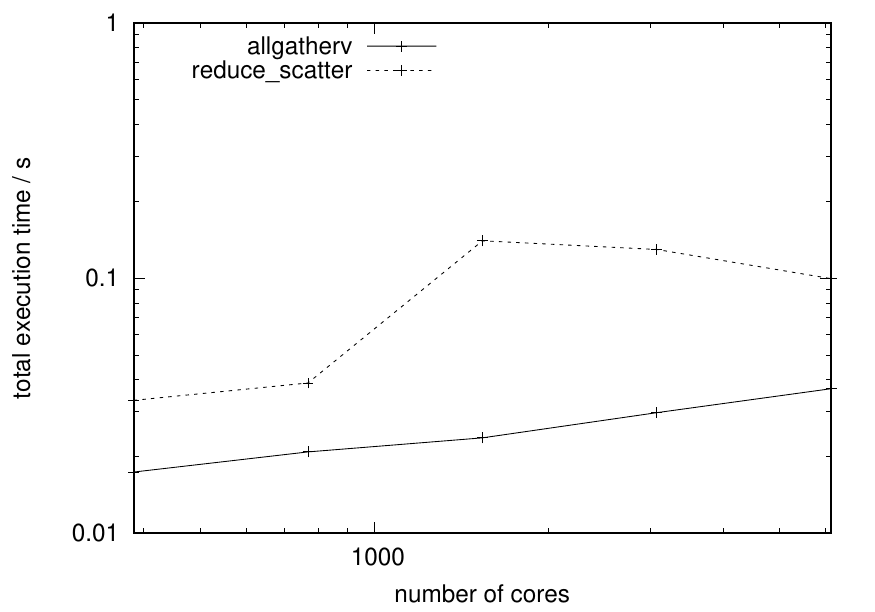}
  \caption{Execution times of collective operations of plasma physics application, rank reordering and reference (left) and without rank reordering (right)}
  \label{fig:plasma}
\end{center}
\end{figure}

Benchmarks are performed with a simplified version of the plasma physics application \cite{ohana_etal:2016}. Parameters are $n_\phi=512$, $n_\theta=1024$, $n_r=512$ in toroidal, poloidal and radial direction, respectively, with $12$ clones and $10^7$ markers. The number of timesteps chosen is $10$ with the size $\Delta t=1$. Apart from the small number of markers this is a typical production run but we note here that the Fourier filter operations apply to grid data only and are therefore independent of the number of markers. Only two Fourier modes are kept in toroidal direction. Thus two messages of 90464 bytes are gathered and distributed to all nodes participating and the reverse is done for the reduction. Figure~\ref{fig:plasma} (left) shows the performance of the {\tt allgatherv} and the {\tt reduce\_scatter} routines in comparison to the Cray MPI reference implementation.
In order to quantify the effect of rank reordering we included graphs (Fig.~\ref{fig:plasma} (right)) for a worst case ordering, messages sorted according to size. For the Cray MPI reference implementation the rank orders are chosen randomly. For up to $128\cdot 12$ cores all benchmarks were done on physical nodes with 12 cores per node, higher core counts were realised on 128 nodes with multiple groups of 12 cores on each node.

\section{Related work}
\label{sec:related_work}
Many efforts were made in order to optimise the collective communication for message passing, especially by exploiting the shared memory on the nodes.
Alm\'asi et al. \cite{almasi_etal:2005} optimised the collectives operations for the BlueGene/L.
\newline Chakraborty et al. \cite{chakraborty_etal:2017} developed MPI collectives using shared memory on the nodes with kernel assistance.
Chan et al. \cite{chan_etal:2007} reimplemented all MPI collective communication routines.
Faraj and Yuan \cite{faraj_yuan:2005} implemented MPI collective communication with an autotuning approach.
Graham and Shipman \cite{graham_shipman2008} optimised shared memory collective communication.
Karwande et al. \cite{karwande_etal:2003} developed a MPI library which selects parameters of the algorithms during compile time of the code.
Patarasuk and Yuan \cite{patarasuk_yuan:2009} proposed a bandwidth optimal allreduce algorithm for SMP clusters.


\section{Conclusions}
\label{sec:conclusions}
In this paper, we optimised the collective communication operations {\tt allgatherv}, {\tt reduce\_scatter} and {\tt allreduce} where we made extensive use of an initialisation phase. The initialisation phase allowed for several optimisations, namely an extensive choice of algorithms, such as recursive multiplication/division, cyclic shifting (Bruck's algorithm) and their derivatives, with different factors (number of ports/substeps) for different steps.
The proper algorithms and their parameters are chosen according to network performance measurements at the installation time of the library.
For {\tt allgatherv} and {\tt reduce\_scatter}, we considered explicitly the occurrence of non-equal message sizes in our algorithms with a rank reordering heuristic. Our {\tt allreduce} for small messages is based on a prime factor decomposition of the number of nodes and {\tt allgather}, for long messages our optimised {\tt allgatherv} and {\tt reduce\_scatter} are consecutively called.

The existing implementation of Cray MPI is outperformed significantly for small and medium message sizes for {\tt allgatherv}. For very large messages the performance is equal between the two options. Although our {\tt reduce\_scatter} is slower than the reference for small messages and a small number of nodes, it outperforms the reference clearly for all other cases. Our {\tt allreduce} is faster than the existing implementation for short messages and long messages. For medium message sizes it shows performance equal to the reference implementation. We mostly outperform MVAPICH with our three routines.

For non-equal message sizes, our routines show additional speedups if the ranks are reordered. This has been shown on the example of a plasma physics application.
The implementation of our algorithms as non-blocking versions is future work.


%
\bibliographystyle{splncs04}
\bibliography{jocksch_parallel19}

\end{document}

%% file: bruck_reverse.inc
  \setlength{\unitlength}{1mm}
  \begin{picture}(170, 65)
    \put(10,60){$n0$}
    \put(20,60){$n1$}
    \put(30,60){$n2$}
    \put(40,60){$n3$}
    \put(50,60){$n4$}
    \put(10,55){$0_0$}
    \put(10,50){$1_0$}
    \put(10,45){$2_0$}
    \put(10,40){$3_0$}
    \put(10,35){$4_0$}
    \put(20,55){$1_1$}
    \put(20,50){$2_1$}
    \put(20,45){$3_1$}
    \put(20,40){$4_1$}
    \put(20,35){$0_1$}
    \put(30,55){$2_2$}
    \put(30,50){$3_2$}
    \put(30,45){$4_2$}
    \put(30,40){$0_2$}
    \put(30,35){$1_2$}
    \put(40,55){$3_3$}
    \put(40,50){$4_3$}
    \put(40,45){$0_3$}
    \put(40,40){$1_3$}
    \put(40,35){$2_3$}
    \put(50,55){$4_4$}
    \put(50,50){$0_4$}
    \put(50,45){$1_4$}
    \put(50,40){$2_4$}
    \put(50,35){$3_4$}
    \put(80,60){$n0$}
    \put(95,60){$n1$}
    \put(110,60){$n2$}
    \put(125,60){$n3$}
    \put(140,60){$n4$}
    \put(80,50){$1_0$}
    \put(80,45){$2_0$}
    \put(80,40){$3_0$}
    \put(95,50){$2_1$}
    \put(95,45){$3_1$}
    \put(95,40){$4_1$}
    \put(110,50){$3_2$}
    \put(110,45){$4_2$}
    \put(110,40){$0_2$}
    \put(125,50){$4_3$}
    \put(125,45){$0_3$}
    \put(125,40){$1_3$}
    \put(140,50){$0_4$}
    \put(140,45){$1_4$}
    \put(140,40){$2_4$}
    \put(76.5,55){$0_0+0_1$}
    \put(91.5,55){$1_1+1_2$}
    \put(106.5,55){$2_2+2_3$}
    \put(121.5,55){$3_3+3_4$}
    \put(136.5,55){$4_4+4_0$}
    \put(21.5,25){$n0$}
    \put(46.5,25){$n1$}
    \put(71.5,25){$n2$}
    \put(96.5,25){$n3$}
    \put(121.5,25){$n4$}
    \put(15,20){$0_0+0_1+0_3$}
    \put(40,20){$1_1+1_2+1_4$}
    \put(65,20){$2_2+2_3+2_0$}
    \put(90,20){$3_3+3_4+3_1$}
    \put(115,20){$4_4+4_0+4_2$}
    \put(19,15){$1_0+1_3$}
    \put(44,15){$2_1+2_4$}
    \put(69,15){$3_2+3_0$}
    \put(94,15){$4_3+4_1$}
    \put(119,15){$0_4+0_2$}
    \put(13.5,5){$n0$}
    \put(48.5,5){$n1$}
    \put(83.5,5){$n2$}
    \put(118.5,5){$n3$}
    \put(153.5,5){$n4$}
    \put(0,0){$0_0+0_1+0_3+0_4+0_2$}
    \put(35,0){$1_1+1_2+1_4+1_0+1_3$}
    \put(70,0){$2_2+2_3+2_0+2_1+2_4$}
    \put(105,0){$3_3+3_4+3_1+3_2+3_0$}
    \put(140,0){$4_4+4_0+4_2+4_3+4_1$}
    \put(60,50){\vector(1,0){8}}
    \put(155,50){\vector(1,0){8}}
    \put(145,20.5){\vector(1,0){8}}
  \end{picture}